\documentclass[11pt]{article}
\usepackage{amsfonts}
\usepackage{mathrsfs}
\usepackage[latin1]{inputenc}
\usepackage{float}
\usepackage{amsmath,amssymb}
\usepackage{latexsym}
\usepackage{epstopdf}
\usepackage{geometry}  
\usepackage[active]{srcltx}
\usepackage{graphicx}
\usepackage{epstopdf}
\usepackage{booktabs}
\usepackage{multirow}
\usepackage{siunitx}
\usepackage{mathtools}
\usepackage{setspace}
\usepackage{listings}
\usepackage{xcolor}
\usepackage{url}
 
\usepackage[
bookmarks=true,         
bookmarksnumbered=true, 
colorlinks=true, pdfstartview=FitV, linkcolor=blue, citecolor=blue,
urlcolor=blue]{hyperref}

\providecommand{\U}[1]{\protect\rule{.1in}{.1in}}

\newcommand{\bi}[1]{\mbox{\boldmath{$ #1 $}}}

\begin{document}
\title{\Large \textbf{\texttt{gcor}}: A Python Implementation of Categorical Gini Correlation and Its Inference}

\vspace{0.5cm}
\author{Sameera Hewage\thanks{CONTACT Sameera Hewage. Email: sameera.hewage@westliberty.edu}}
\date{%
   Department of Physical Sciences \& Mathematics, West Liberty University, West Liberty, WV 26074, USA.\\[1em] 
    \today
}

\maketitle
\begin{abstract}
Categorical Gini Correlation (CGC) proposed by Dang et al. \cite{Dang2021} measures the dependence between a numerical variable and a categorical variable. It has appealing properties compared to existing dependence measures, such as zero correlation mutually implying independence between the variables. It has also shown superior performance over existing methods when applied to feature screening for classification. This article presents a Python implementation for computing CGC, constructing confidence intervals, and performing independence tests based on it. Efficient algorithms have been implemented for all procedures, and they have been optimized using vectorization and parallelization to enhance computational efficiency.

\end{abstract}

\noindent 

\section{Introduction}
Categorical Gini correlation (CGC), also known as \textit{Gini distance correlation}, was proposed by Dang \textit{et al.}~\cite{Dang2021} to measure the dependence between a continuous random vector and a categorical variable. CGC has been shown to possess several desirable properties compared to many existing dependence measures. Inference procedures for CGC have been developed in both fixed-dimensional~\cite{Dang2021, Hewage, Sang2024} and high-dimensional~\cite{Sang2023} settings. Moreover, CGC has been employed as a dependence measure in recent feature selection methods for classification, including those proposed in~\cite{Zhang2019, Sang2023, Sang2024, Shang2023}.

Let \(\bi{X}\) be a continuous random vector following the distribution \(F\) in \(\mathbb{R}^d\). Let \(Y\) be a categorical response variable with possible values \(L_1, \dots, L_K\) and a distribution \(P_Y\) such that \(P(Y = L_k) = p_k > 0\) for \(k = 1, 2, \dots, K\). Assume that the conditional distribution of \(\bi{X}\) given \(Y = L_k\) is \(F_k\). Then Gini covariance   and correlation are defined 
in \cite{Dang2021} as
\[
\text{gCov}(\bi{X}, Y) = \Delta - \sum_{k=1}^K p_k \Delta_k,
\]
and
\begin{align}\label{gc}
\rho_g(\bi{X}, Y) = \frac{\Delta - \sum_{k=1}^K p_k \Delta_k}{\Delta},
\end{align}
where \begin{align*}
\Delta = \mathbb{E} \|\bi{X}_1 - \bi{X}_2\|, \quad \Delta_k = \mathbb{E} \|\bi{X}_1^{(k)} - \bi{X}_2^{(k)}\|
\end{align*} are the multivariate Gini mean differences \cite{Yitzhaki13, HewageDiss, HewageGMD} for \( F \) and \( F_k \) respectively with
\( (\bi{X}_1, \bi{X}_2)^T \) and \( (\bi{X}_1^{(k)}, \bi{X}_2^{(k)})^T \) representing independent pair variables each drawn independently from \( F \) and \( F_k \), respectively.
Thus, it is observed that the Gini correlation can be interpreted as the ratio of between variation and overall variation analogous to Pearson $R^2$ in ANOVA model \cite{Dang2021}. 

The categorical Gini covariance measures dependence by assessing the weighted distance between the marginal and conditional distributions. Let $\psi_k$ and $\psi$ be the charactersitic functions  of  $F_k$ and $F$, respectively. In fact, the Gini covariance in (\ref{gc}) can be defined by 
\begin{equation}\label{Gcov}
\mbox{gCov}(\bi X,Y) = c(d) \sum_{k=1}^K p_k \int_{\mathbb{R}^d} \frac{|\psi_k(\bi t) -\psi(\bi t)|^2}{\|\bi t\|^{d+1}} d\bi t,
\end{equation}
where $c(d) = \Gamma((d+1)/2)/\pi^{(d+1)/2}$. 
We observe that \( \text{gCov}(\bi{X}, Y) \geq 0 \), and \( \text{gCov}(\bi{X}, Y) = 0 \) if and only if \( \bi{X} \) and \( Y \) are independent
 \cite{Dang2021}. The associated Gini correlation standardizes this covariance to ensure it falls within the interval \([0, 1]\).

The distance correlation introduced by Sz\'ekely, Rizzo, and Bakirov~\cite{Szekely2007, Szekely2009, Szekely2013a, Ramos} is a widely used dependence measure capable of assessing the association between a continuous random vector and a categorical variable. In contrast, CGC has been shown to offer several advantages: (a) improved computational efficiency, (b) simplified statistical inference, and (c) greater robustness when handling unbalanced data. These appealing properties motivate the development of a Python implementation of CGC and its associated inference procedures, particularly given Python's growing popularity for data analysis and statistical computing. Nguyen and Dang have implemented CGC and its inference procedures in the R package \texttt{GiniDistance}~\cite{Dao}. However, to the best of our knowledge, no Python implementation currently exists.

The remainder of the paper is organized as follows. In Section~\ref{sec:cgc}, we review some results on CGC, introduce three Python functions, and illustrate each with a real data example. Section~\ref{sec:impact} presents the impact of the proposed implementation and its applications. In Section~\ref{sec:conclusion}, we provide concluding remarks and discuss future work.

\section{Functionalities of the package}\label{sec:cgc}

In this section, we first review key results on CGC and its related inference methods, and finally illustrate the Python implementation using a real dataset.

\subsection{Estimation of CGC}
Consider a sample \( \mathcal{D} = \{ (\bi{X}_1, Y_1), (\bi{X}_2, Y_2), \dots, (\bi{X}_n, Y_n) \} \) drawn from the joint distribution of \( \bi{X} \) and \( Y \). We can decompose \( \mathcal{D} \) as 
\[
\mathcal{D} = \mathcal{D}_1 \cup \mathcal{D}_2 \cup \dots \cup \mathcal{D}_K,
\]
where 
\[
\mathcal{D}_k = \{\bi{X}_1^{(k)}, \bi{X}_2^{(k)}, \dots, \bi{X}_{n_k}^{(k)}\}
\]
denotes the sample with \( Y_i = L_k \), and \( n_k \) is the number of samples in the \( k^{\text{th}} \) class.

Categorical Gini correlation can then be estimated unbiasedly as a function of \( U \)-statistics \cite{Dang2021}:
\[
\hat{\rho}_g(\bi{X}, Y) = \frac{\tilde{U} - \sum_{k=1}^K \hat{p}_k \tilde{U}_k}{\tilde{U}},
\]
where \( \hat{p}_k = \frac{n_k}{n} \), and
\begin{align*} \label{Ustat}
\tilde{U}_k &= {n_k\choose 2}^{-1} \sum_{1 \leq i,j \leq n_k} \|\bi{X}_i^{(k)} - \mathbf{X}_j^{(k)}\|, \\
\tilde{U} &= {n\choose 2}^{-1} \sum_{1 \leq i,j \leq n} \|\bi{X}_i - \mathbf{X}_j\|.
\end{align*}

In this work, the Python function \texttt{gcor} is introduced to calculate the CGC based on the above estimation.

\subsection{Confidence Interval}
Dang et al.~\cite{Dang2021} established that when \( \bi{X} \) and \( Y \) are dependent, the estimator \(\hat{\rho}_g(\bi{X}, Y)\) satisfies the following asymptotic normality:

\[
\sqrt{n} \left( \hat{\rho}_g(\bi{X}, Y) - \rho_g(\bi{X}, Y) \right) \xrightarrow{d} \mathcal{N}(0, \sigma_g^2),
\]
where \( \sigma_g^2 \) is the asymptotic variance given by \cite{Dang2021}.

Confidence intervals for CGC can be constructed based on this asymptotic normality. However, the variance \(\sigma_g^2\) is often difficult to compute directly due to its complex form. To address this, one can estimate \(\sigma_g^2\) by employing the jackknife method.

Let $\hat {\rho_g}_{(-i)}$ be the jackknife pseudo value of the Gini correlation estimator $\hat {\rho_g}$ based on the sample with the $i^{th}$ observation deleted. Then, the jackknife estimator of $\sigma^2_g$ is
\begin{align}\label{jel_var}
\widehat{\sigma}_g^2 =\dfrac{n-1}{n}\sum_{i=1}^n (\hat{\rho}_{g(-i)} -\bar{\hat{\rho}}_{g(\cdot)})^2,
\end{align}
where $\bar{\hat{\rho}}_{g(\cdot)} =\dfrac{1}{n} \sum_{i=1}^n {\hat{\rho}}_{g(-i)},$ see \cite{Shao96}. 
An approximate \((1-\alpha) \times 100\%\) confidence interval for the categorical Gini Correlation can then be constructed as
\[
\left[ \hat{\rho}_g(\bi{X}, Y) - z_{\alpha/2} \frac{\hat{\sigma}_g}{\sqrt{n}}, \quad \hat{\rho}_g(\bi{X}, Y) + z_{\alpha/2} \frac{\hat{\sigma}_g}{\sqrt{n}} \right],
\]
where \( z_{\alpha/2} \) is the upper \(\alpha/2\) quantile of the standard normal distribution. The Python function \texttt{gcorCI} is introduced to calculate the confidence interval in the present work. 

\subsection{Independence Test}
The independence test based on CGC is stated as
\begin{align}\label{indtest}
H_0: \rho_g(\mathcal{\bi{X}}, Y) = 0 \quad \text{vs} \quad H_1: \rho_g(\mathcal{\bi{X}}, Y) > 0.
\end{align}

The null hypothesis in test (\ref{indtest}) is equivalent to the null hypothesis of the \(K\)-sample test : 
\[
\mathcal{H}_0': F_1 = F_2 = \cdots = F_K = F.
\]
In other words, this means that the distributions for each category, \(F_1, F_2, \ldots, F_K\), are identical. We reject \( \mathcal{H}_0 \) or \( \mathcal{H}_0' \) when the observed value of \( \hat{\rho}_g \) is sufficiently large. Calculating the critical value for the test is challenging due to its dependence on unknown distribution parameters and the complex mixture distribution of the test statistic. To address this, as suggested by \cite{Dang2021}, a permutation procedure is employed to estimate both the critical value and the \( p \)-value. We introduce the Python function \texttt{independence\_test} to perform the independence test.

\subsection{Generalized Categorical Gini Correlation}
For a nondegenerate random vector $\bi X$ in $\mathbb{R}^d$ and a categorical variable $Y$, if $E[||\bi X||^\alpha] < \infty$ for $\alpha \in (0, 2)$, the generalized Gini correlation between $\bi X$ and $Y$ is defined as:
$$ \rho_g(\bi X, Y; \alpha) =  \frac{\Delta(\alpha) - \sum_{k=1}^K p_k \Delta_k(\alpha)}{\Delta(\alpha)}. $$
According to \cite{Dang2021} , a computational consideration is the choice of $\alpha$, which is the parameter for the distance metric in $\mathbb{R}^d$. By their recommendation, a natural choice is $\alpha = 1$, corresponding to the Euclidean distance, which facilitates fast algorithms for the univariate case. However, in the presence of outliers, a smaller $\alpha$ value is preferred to ensure that CGC remain insensitive to these outliers. We provide a parameter called ``\texttt{alpha}'' in each function to accommodate the generalized CGC.
%


\subsection{Illustrative example}

To demonstrate the applicability of the proposed CGC functions, we use the well-known IRIS dataset, which is widely used in machine learning and statistics. Originally introduced by the British biologist and statistician Ronald A. Fisher in 1936 for discriminant analysis, the dataset comprises 150 samples of iris flowers from three species: Setosa, Versicolor, and Virginica. Each flower is described by four numerical features: sepal length, sepal width, petal length, and petal width. The target variable is the species classification.

We begin by illustrating how to compute the CGC using the \texttt{gcor} function. The first example evaluates the correlation between a single numerical feature (sepal length) and the species label, while the second example applies the method to a multivariate case using both sepal length and width.

\definecolor{lightgray}{gray}{0.95}
\begin{lstlisting}[language=Python, 
  backgroundcolor=\color{lightgray}, 
  basicstyle=\footnotesize\ttfamily, 
  frame=single, 
  caption=Usage of \texttt{gcor} function with a univariate numerical variable,
  label={lst:gcor1}
]
from sklearn.datasets import load_iris

iris = load_iris()
x = iris.data[:, 0]  # Sepal length 
y = iris.target       # Species

x = gcor(x, y, alpha=1)

Output: Categorical Gini Correlation: 0.397830
\end{lstlisting}

Listing~\ref{lst:gcor1} loads the Iris dataset and selects sepal length as the predictor variable \( X \). The \texttt{gcor} function then computes the CGC between sepal length and the species labels \( Y \).

\begin{lstlisting}[language=Python, 
  backgroundcolor=\color{lightgray}, 
  basicstyle=\footnotesize\ttfamily, 
  frame=single, 
  caption=Usage of \texttt{gcor} function with multivariate numerical variables,
  label={lst:gcor2}
]
from sklearn.datasets import load_iris 
iris = load_iris()

x = iris.data[:, :2]  # First two features: sepal length and width
y = iris.target

gcor(x, y, alpha=1)

Output: Categorical Gini Correlation: 0.357026
\end{lstlisting}

In Listing~\ref{lst:gcor2}, both sepal length and width are used as predictor variables. The \texttt{gcor} function computes the CGC between the multivariate predictor \( \bi{X} \) and the species labels \( Y \).

\begin{lstlisting}[language=Python, 
  backgroundcolor=\color{lightgray}, 
  basicstyle=\footnotesize\ttfamily, 
  frame=single, 
  caption=Usage of \texttt{gcorCI} function for confidence interval estimation,
  label={lst:gcor3} 
]
from sklearn.datasets import load_iris 
iris = load_iris()

x = iris.data[:, :2]  # First two features: sepal length and width
y = iris.target

gcorCI(x, y, clevel=0.95)

Output: 95% Confidence Interval: [0.306404, 0.407647]
\end{lstlisting}

Listing~\ref{lst:gcor3} shows how to compute an approximate 95\% confidence interval for the CGC using the \texttt{gcorCI} function. Here, the predictors \( \bi{X} \) are again sepal length and width, and \( Y \) denotes the species labels.

\begin{lstlisting}[language=Python, 
  backgroundcolor=\color{lightgray}, 
  basicstyle=\footnotesize\ttfamily, 
  frame=single, 
  caption=Usage of \texttt{independence\_test} function for testing independence,
  label={lst:gcor4}
]
np.random.seed(123)

n_per_group = 50
x1 = np.random.normal(loc=0, scale=1, size=(n_per_group, 2))
x2 = np.random.normal(loc=0, scale=1, size=(n_per_group, 2))
x3 = np.random.normal(loc=0, scale=1, size=(n_per_group, 2))

x = np.vstack([x1, x2, x3])
y = np.array([0]*n_per_group + [1]*n_per_group + [2]*n_per_group)

p_value, reject_null = independence_test(x, y, B=1000)

Output: P-value: 0.6100
Fail to reject null hypothesis.
\end{lstlisting}

Listing~\ref{lst:gcor4} demonstrates the use of the \texttt{independence\_test} function. It simulates three independent groups, each with 50 samples from the same bivariate normal distribution. The predictor matrix \( \bi{X} \) and group labels \( Y \) are passed to the independence test function, which uses CGC and a permutation test with 1000 iterations. The resulting p-value of 0.6100 suggests that there is no significant dependence between the features and the group labels, which aligns with the data generation setup.

\subsection{Implementation Details}
The implementation of the CGC functions leverages vectorized computations to enhance efficiency. Parallelization has been applied wherever possible, such as in the computation of the jackknife variance estimator for confidence interval calculation, and also during permutation tests, which involve repeated evaluation of the correlation measure. Since the pairwise distances between samples remain fixed across permutations while only the labels are shuffled, the distance matrix needs to be computed just once, thereby reducing redundant computations. This approach enables straightforward use of parallel processing tools, such as the \texttt{joblib} or \texttt{multiprocessing} libraries in Python, to perform permutations concurrently. These computations are naturally suited for parallel execution since each task is independent, and the overall speed improvement generally increases with the number of CPU cores used which makes it feasible to handle large datasets efficiently.

\subsection{Computational Performance}\label{sec:benchmark}

To assess the computational performance of \texttt{gcor}, we compared its runtime with the R package \texttt{GiniDistance}~\cite{Dao} over a range of sample sizes $n \in \{100, 300, 500, 1000, 2000, 3000\}$ and dimensions $d \in \{1, 2, 5, 10, 20\}$. All experiments used $K = 3$ balanced classes. The Python \texttt{gcor} function was benchmarked against the R \texttt{gCor} implementation from \texttt{GiniDistance}, ensuring that both implementations computed the same underlying statistic. Independent data were generated in each experiment, and the reported runtime corresponds to the median over five replications. All computations were carried out on a machine with an AMD Ryzen 7 5800H processor (3.20\,GHz) and 16\,GB RAM running 64-bit Windows~11, using Python~3.11 and R~4.4.1.

For the univariate setting ($d = 1$), the Python implementation uses an $O(n \log n)$ algorithm based on the identity
\[
\sum_{i < j} |x_i - x_j|
=
\sum_{j=1}^n (2j - n - 1)\, x_{(j)},
\]
where $x_{(j)}$ denotes the $j$-th order statistic. This approach avoids constructing the full $O(n^2)$ pairwise distance matrix and is consistent with the computational strategy used internally by the R package for univariate data \cite{Dang2021}. For higher-dimensional settings ($d \geq 2$) or when $\alpha \neq 1$, both implementations use the general $O(n^2)$ pairwise distance computation.

\begin{figure}[H]
    \centering
    \includegraphics[width=0.72\linewidth]{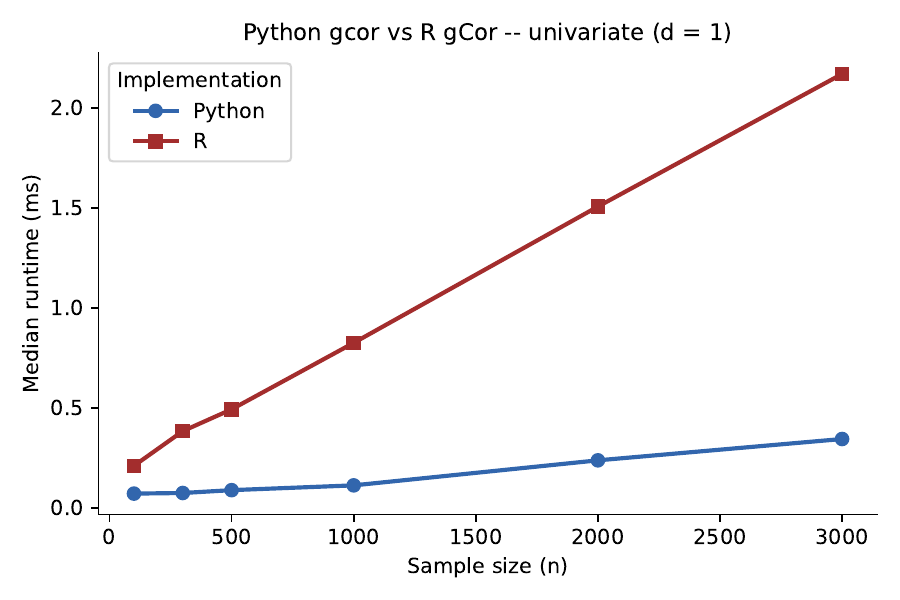}
    \caption{Median runtime (ms) versus sample size $n$ for the univariate setting ($d = 1$), comparing the Python \texttt{gcor} implementation and the R \texttt{GiniDistance} package (\texttt{gCor}). Both implementations use an $O(n \log n)$ algorithm for $d = 1$ and $\alpha = 1$.}
    \label{fig:bench_uni}
\end{figure}

Figure~\ref{fig:bench_uni} presents the runtime comparison for the univariate case. At $n = 1000$, the Python implementation required $0.11$\,ms compared with $0.83$\,ms in R, while at $n = 3000$ the runtimes were $0.34$\,ms and $2.17$\,ms, respectively. These correspond to speedups of approximately $7.4\times$ and $6.3\times$. Similar behavior is observed across all sample sizes considered, indicating that the vectorized NumPy implementation of the $O(n \log n)$ algorithm provides substantial computational gains in the univariate setting.

\begin{figure}[H]
    \centering
    \includegraphics[width=\linewidth]{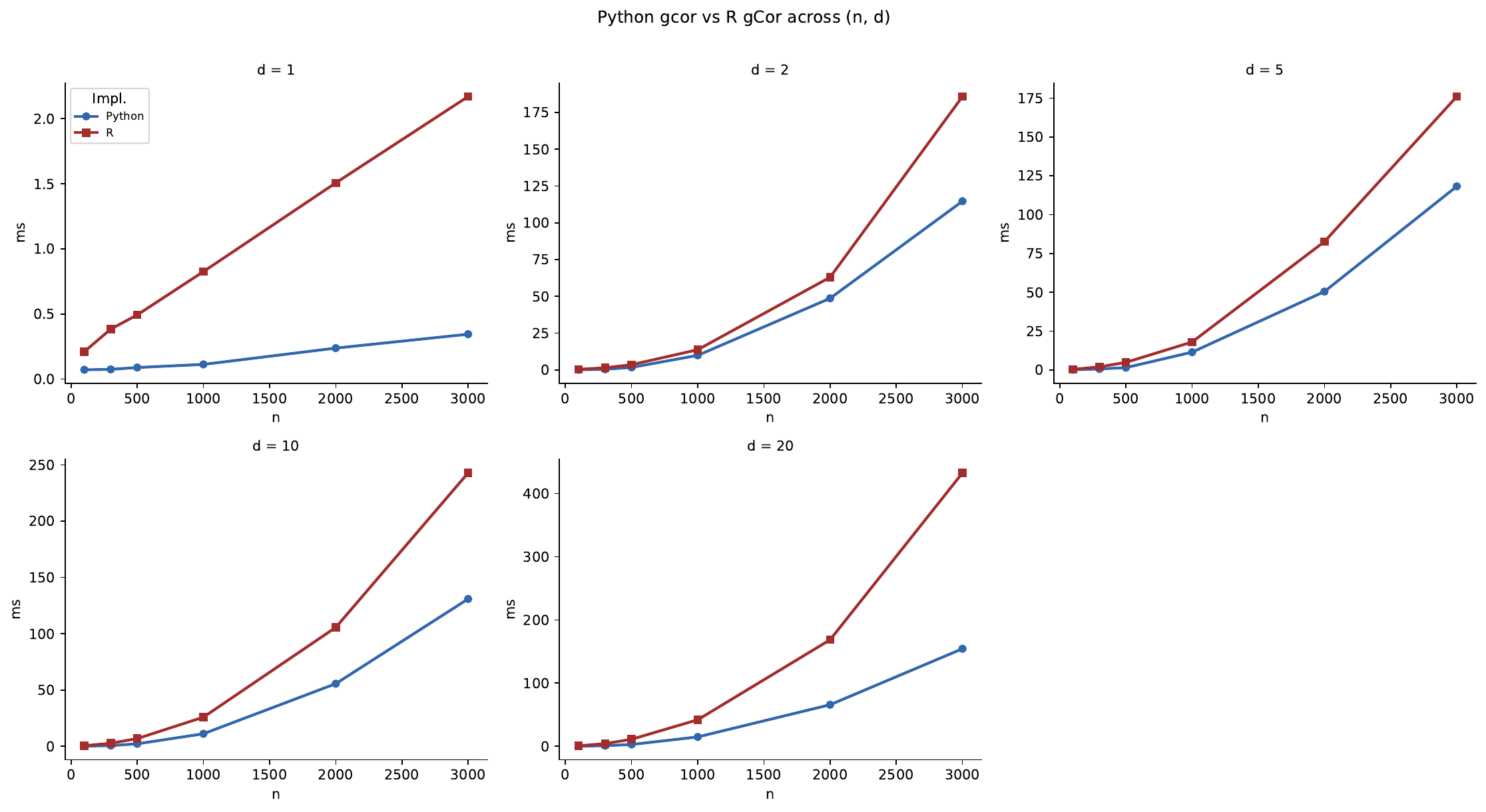}
    \caption{Median runtime (ms) versus sample size $n$ across dimensions $d \in \{1, 2, 5, 10, 20\}$ for both implementations. Each panel corresponds to a fixed dimension.}
    \label{fig:bench_dims}
\end{figure}

Figure~\ref{fig:bench_dims} extends the comparison to higher-dimensional settings. For $d \geq 2$, both implementations rely on the general $O(n^2)$ pairwise algorithm. Across the tested configurations, the Python implementation consistently achieved lower runtimes, with observed speedups ranging from approximately $1.3\times$ to $4.8\times$. For example, when $n = 3000$, the Python implementation required $114.6$\,ms for $d = 2$ and $154.3$\,ms for $d = 20$, whereas the corresponding runtimes in R were $185.7$\,ms and $432.6$\,ms. The effect of increasing dimension on runtime was less pronounced in Python than in R. At $n = 3000$, increasing the dimension from $d = 2$ to $d = 20$ increased runtime by a factor of approximately $1.35$ in Python compared with $2.33$ in R. This suggests that the vectorized Python implementation scales efficiently as dimensionality increases.

\begin{figure}[H]
    \centering
    \includegraphics[width=0.72\linewidth]{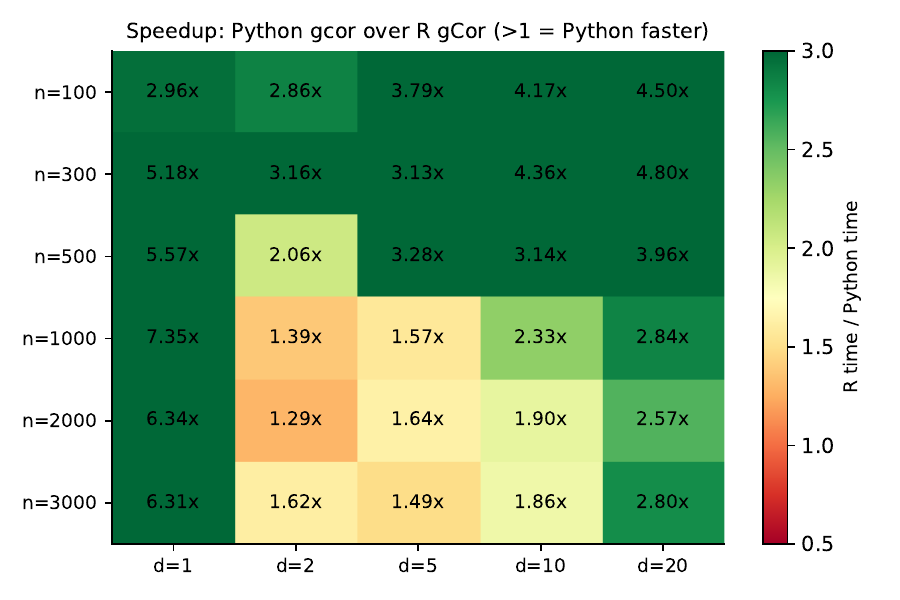}
    \caption{Speedup of the Python implementation relative to the R implementation (\texttt{gCor}), computed as the ratio of median R runtime to median Python runtime across the $(n,d)$ grid. Values greater than 1 indicate faster execution in Python.}
    \label{fig:bench_heatmap}
\end{figure}

The relative speedup of the Python implementation over the R implementation is summarized in Figure~\ref{fig:bench_heatmap}. Across the entire $(n,d)$ grid, Python achieved lower runtimes than R, with speedups ranging from approximately $1.3\times$ at $(n = 2000, d = 2)$ to $7.4\times$ at $(n = 1000, d = 1)$. The largest gains occur in the univariate setting, where the optimized $O(n \log n)$ algorithm is most effective, as well as in higher-dimensional settings where vectorized numerical operations become increasingly advantageous.

\begin{figure}[H]
    \centering
    \includegraphics[width=0.65\linewidth]{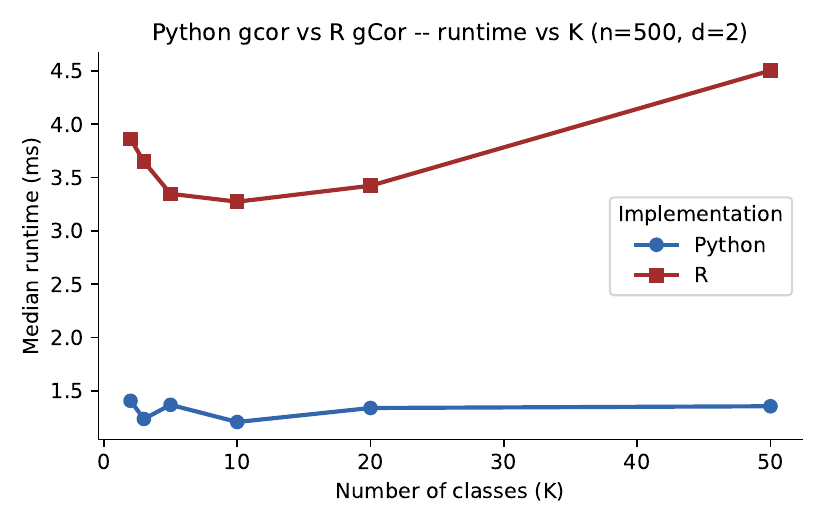}
    \caption{Median runtime (ms) versus number of classes $K$ for both implementations with $n = 500$ and $d = 2$. Runtime remains relatively stable as $K$ increases, indicating that pairwise distance computation dominates the overall computational cost.}
    \label{fig:bench_k}
\end{figure}

Figure~\ref{fig:bench_k} examines the effect of the number of classes $K$ on runtime while holding $n = 500$ and $d = 2$ fixed. The Python runtimes remained relatively stable, ranging from $1.21$\,ms to $1.41$\,ms across $K \in \{2,3,5,10,20,50\}$. The corresponding R runtimes ranged from $3.27$\,ms to $4.50$\,ms. For both implementations, runtime was only weakly affected by the number of classes, indicating that the dominant computational cost arises from pairwise distance calculations rather than the class-specific aggregation step.

\subsection{Reproducibility}

All code developed in this study for computing the CGC, constructing confidence intervals, and performing independence tests is publicly available at
\url{https://github.com/sameera-hewage/gcor}.
\section{Impact and applications}\label{sec:impact}

In recent years, methods based on categorical Gini correlation have attracted growing interest for their ability to capture complex dependence structures between numerical and categorical variables. This work presents a unified and efficient Python framework for computing categorical Gini correlation and conducting related inference, aiming to make these tools more accessible to researchers and practitioners.

Categorical Gini correlation has already found applications in machine learning, particularly in feature selection for classification tasks \cite{Zhang2019, Sang2023, Sang2024, Shang2023}. Its utility is especially evident in high-dimensional settings, where computational efficiency becomes critical. The Python implementation developed here focuses on improving performance through vectorization and parallelization, enabling faster execution in practical data analysis scenarios. This framework is well-suited for use in a variety of fields, including bioinformatics, social sciences, and applied machine learning, where analyzing relationships between mixed data types is a key concern.

Dependence measures are increasingly being incorporated into modern statistical learning frameworks involving feature selection, imbalanced learning, risk prediction, and dependence modeling. Recent studies have explored supervised dependence-based feature selection for disease prediction \cite{Aich1}, synthetic data generation methods for imbalanced classification \cite{Aich2}, fusion of clinical and genomic machine learning risk scores \cite{Aich3}, and Bayesian approaches for modeling complex tail dependence structures \cite{Aich4}. Related developments have also considered copula-driven neural network frameworks for capturing non-Gaussian spatial dependence \cite{Aich5}. Since these methodologies rely heavily on accurately characterizing nonlinear dependence structures, tools such as the \texttt{gcor} package may provide useful computational support in machine learning and statistical inference applications involving mixed data types.


\section{Conclusions}\label{sec:conclusion}

This article introduces a Python implementation of the categorical Gini correlation, providing a comprehensive and efficient toolkit for measuring dependence between numerical and categorical variables. Beyond including core statistical measures and inference procedures, the framework has been developed with a strong emphasis on computational efficiency and flexibility. Key design features include vectorized operations and potential for parallelization, which contribute to improved performance compared to existing alternatives.

The growing interest in categorical Gini correlation and related methods is reflected in their adoption across novel applicatoins including statistics \cite{suresh, Gamero} and machine learning \cite{Wang2025}. This implementation aims to make these advanced statistical tools more accessible to researchers and practitioners, supporting tasks like feature selection and association analysis in complex data scenarios. Future work will focus on expanding the functionality, enhancing documentation with practical examples, and integrating further theoretical developments to strengthen the package.

We remain dedicated to supporting the open-source community. Feedback from users plays a key role in guiding ongoing enhancements. We also welcome contributions from others to help expand and refine the categorical Gini correlation toolkit collaboratively.


%
%

\end{document}